\documentclass[prd,reprint,amsmath,amssymb,floatfix,aps,longbibliography,superscriptaddress,nofootinbib]{revtex4-2}

\usepackage{amsmath,graphicx,amssymb,xcolor,booktabs,braket,multirow,cancel,enumerate,enumitem}
\usepackage[colorlinks=true, allcolors=blue]{hyperref}
\usepackage[normalem]{ulem}
\usepackage{soul}
\usepackage{ytableau}

\setstcolor{red}

\begin{document}

\title{
High Reheating Temperature without Axion Domain Walls
}

\author{Shota Nakagawa}
\email{shota.nakagawa@sjtu.edu.cn}
\affiliation{Tsung-Dao Lee Institute, Shanghai Jiao Tong University, \\ No.~1 Lisuo Road, Pudong New Area, Shanghai 201210, China}
\affiliation{School of Physics and Astronomy, Shanghai Jiao Tong University, \\ 800 Dongchuan Road, Shanghai 200240, China}

\author{Yuichiro Nakai}
\email{ynakai@sjtu.edu.cn}
\affiliation{Tsung-Dao Lee Institute, Shanghai Jiao Tong University, \\ No.~1 Lisuo Road, Pudong New Area, Shanghai 201210, China}
\affiliation{School of Physics and Astronomy, Shanghai Jiao Tong University, \\ 800 Dongchuan Road, Shanghai 200240, China}

\author{Yu-Cheng Qiu}
\email{ethan.qiu@cityu.edu.hk}
\affiliation{Department of Physics, City University of Hong Kong, Kowloon, Hong Kong SAR, China}

\author{Lingyun Wang}
\email{lyunwang@whu.edu.cn}
\affiliation{Department of Physics, School of Physics and Technology, Wuhan University, No.~299 Bayi Road, Wuhan 430072, China}

\author{Yaoduo Wang}
\email{yaoduowang@sjtu.edu.cn}
\affiliation{Tsung-Dao Lee Institute, Shanghai Jiao Tong University, \\ No.~1 Lisuo Road, Pudong New Area, Shanghai 201210, China}
\affiliation{School of Physics and Astronomy, Shanghai Jiao Tong University, \\ 800 Dongchuan Road, Shanghai 200240, China}

\date{\today}

\begin{abstract}

We investigate a cosmological scenario in which the Peccei-Quinn (PQ) symmetry remains broken in the entire history of the Universe, thereby avoiding the formation of axion strings and domain walls. Contrary to the conventional expectation, it is demonstrated that appropriately chosen scalar interactions are able to keep the PQ symmetry broken at arbitrarily high temperatures. We carefully examine the finite-temperature effective potential in a model with two PQ breaking scalar fields. The existence of flat directions  plays a vital role in suppressing axion isocurvature perturbations during inflation by stabilizing a PQ field at a large field value. The viable parameter space consistent with theoretical and observational constraints is identified. Our scenario provides a minimal path for PQ symmetry breaking that addresses both the axion domain wall and isocurvature problems while permitting arbitrarily high reheating temperatures accommodating high-scale baryogenesis scenarios such as thermal leptogenesis.

\end{abstract}

\maketitle

\section{Introduction}

It has been considered that the Universe goes through a reheating epoch after inflation where the inflaton decays and creates the thermal bath of the Standard Model (SM) particles. 
Assuming the instantaneous decay after inflation,~\footnote{
The reheating temperature is different from the maximum temperature in the Universe.
The latter is important for our motivation or non-restoration of symmetry.
In fact, the reheating process may last long, especially if the interaction of the inflaton with the SM particles is small which is expected for suppressed corrections to the inflaton potential~\cite{Chung:1998rq,Davidson:2000er,Allahverdi:2002pu,Harigaya:2013vwa,Mukaida:2015ria}.
Nevertheless, we simply assume the reheating temperature is the maximum temperature throughout the paper.} 
the energy density of the inflaton is transferred into that of the SM thermal bath, and then the reheating temperature is naively estimated as $T_{\rm R} \sim \sqrt{H_{\rm inf} M_{\rm Pl}}$, where $H_{\rm inf}$ is the Hubble scale during inflation and $M_{\rm Pl} \simeq 2.4\times 10^{18} \, \rm GeV$ denotes the reduced Planck scale. 
For instance, $H_{\rm inf}\sim 10^{13} \, \rm GeV$ leads to $T_{\rm R} \sim 10^{15} \, \rm GeV$, which is close to the scale of the grand unified theory (GUT).
With such a high reheating temperature,  high-scale baryogenesis scenarios such as leptogenesis~\cite{Fukugita:1986hr} work well without complicated assumptions.
A high reheating temperature usually restores a broken symmetry~\cite{Kirzhnits:1972iw,Kirzhnits:1972ut,Dolan:1973qd,Weinberg:1974hy,Kirzhnits:1976ts}, and when the symmetry breaking takes place, associated topological defects may be produced. 

The Lagrangian of quantum chromodynamics (QCD) allows a CP-violating parameter $\bar \theta$, which is a combination of the Yang-Mills vacuum parameter and a complex phase of the determinant of quark Yukawa matrices. 
Following the naturalness principle, these two uncorrelated components shall result in $\bar \theta = \mathcal O(1)$.
However, experimental bounds on the neutron electric dipole moment (EDM) indicate $\bar \theta < 10^{-10}$~\cite{Baker:2006ts,Pendlebury:2015lrz,Abel:2020pzs}.
This is the strong CP problem.
The most common approach to the problem is introducing the QCD axion, associated with spontaneous breaking of a global Peccei-Quinn (PQ) symmetry~\cite{Peccei:1977hh,Peccei:1977ur,Weinberg:1977ma,Wilczek:1977pj}. 
The axion degenerates with $\bar \theta$, making it dynamical and stabilized at $\bar \theta \to 0$ due to its potential from QCD non-perturbative effects. 

The axion decay constant $f_a$ labels the scale of the PQ symmetry breaking and the strength of interactions between the axion and SM particles. 
A smaller $f_a$ indicates stronger interactions, which are translated to higher detectability (see e.g. Refs.~\cite{Irastorza:2018dyq,Irastorza:2021tdu,Caputo:2024oqc} for reviews on axion detection).
Cosmological consequences of the QCD axion are different, depending on $f_a$ and the reheating temperature $T_{\rm R}$.
For $f_a \gg T_{\rm R}$, the PQ symmetry is never thermally restored during the thermal history of the Universe.
During inflation, the axion obtains quantum fluctuations, $\langle \delta a^2 \rangle \sim (H_{\rm inf}/2\pi)^2$. 
When the axion dark matter is produced via the misalignment mechanism~\cite{Preskill:1982cy,Abbott:1982af,Dine:1982ah}, it has the isocurvature density perturbations, stringently constrained by observations of Cosmic Microwave Background (CMB) anisotropies. 
The power spectrum of this isocurvature perturbation is given by
\begin{equation}
    \mathcal P_{S_c} = \left( \frac{\delta \Omega_a}{\Omega_c} \right)^2 \simeq 1.4 \times \theta_{i,\rm eff}^2 \left(\frac{H_{\rm inf}}{2\pi f_{\rm inf}} \right)^2 \left( \frac{f_a}{10^{12}\,{\rm GeV}} \right)^{2.38}\;,
    \label{isocurvature}
\end{equation}
where $\Omega_c h^2 \simeq 0.12$~\cite{Planck:2018vyg} denotes the observed density parameter of cold dark matter with $h$ the reduced Hubble parameter, $\delta \Omega_a$ is that of the axion fluctuations, and we assume the quadratic potential for the axion in estimating the axion abundance.
The effective misalignment angle is $\theta_{i,\rm eff}^2 = \theta_i^2 + (H_{\rm inf}/2\pi f_{\rm inf})^2$,
where $\theta_i$ is the intrinsic misaligned angle and $f_{\rm inf}$ denotes the axion decay constant during inflation.
Uncorrelated isocurvature perturbations of cold dark matter are constrained as~\cite{Planck:2018jri} 
\begin{equation}
    \alpha_c \equiv \frac{\mathcal P_{S_c}}{\mathcal P_\zeta + \mathcal P_{S_c}} < 0.033\;,
    \label{eq:alpha}
\end{equation}
where $\mathcal P_\zeta \simeq 2 \times 10^{-9}$ is the power spectrum of the curvature perturbations.
If $f_{\rm inf} \simeq f_a$, then the inflation scale is severely constrained. For example, $f_a = 10^{12} \, \rm GeV$ indicates that $H_{\rm inf} \lesssim 4.4\times 10^{8} \, \rm GeV$ under $\theta_i = 0.1$.
Such a low-scale inflation cannot avoid unnatural fine-tuning of the potential and the initial condition.
Moreover, high-scale baryogenesis scenarios do not favor low reheating temperatures.

For $f_a \ll T_{\rm R}$, it has been conventionally expected that the PQ symmetry is restored after inflation.
Then, its spontaneous breaking during the thermal history of the Universe leads to the formation of associated axion strings.
When the so-called domain wall number $N_{\rm DW}$ is larger than one, a stable axion string-domain wall network is formed at around the QCD phase transition.
In the scaling regime, the energy density of the domain wall scales as $\propto R^{-2}$, where $R$ is the scale factor of the Friedmann‐Lemaitre‐Robertson-Walker (FLRW) metric, which means that it soon dominates the Universe, causing a serious cosmological difficulty~\cite{Zeldovich:1974uw,Vilenkin:1984ib}.
A common approach to this domain wall problem is to introduce an explicit PQ breaking bias term~\cite{Vilenkin:1981zs,Sikivie:1982qv,Gelmini:1988sf}, that lifts the degeneracy between vacua and destabilizes the string-domain wall network. 
However, such a bias term must be finely tuned so that its contribution to the axion potential aligns its minimum with that of the QCD-induced potential; otherwise, the original solution to the strong CP problem is invalidated.~\footnote{
For recent solutions to this issue, see e.g. Refs.~\cite{Girmohanta:2024nyf,Hor:2025gxo,Hao:2025kcz}.}

An alternative solution to the axion domain wall problem was proposed by the authors of Refs.~\cite{Dvali:1995cc,Dvali:1996zr}: they discussed the idea that the PQ symmetry is not restored even at high temperatures, $T \gg f_a$.
Since the PQ symmetry is broken in the entire history of the Universe, the axion strings and domain walls are not formed.
Such a non-restoration of the PQ symmetry could be realized by introducing additional scalar fields coupling with a PQ field responsible for the spontaneous PQ symmetry breaking.
While the idea is attractive, the authors of Refs.~\cite{Dvali:1995cc,Dvali:1996zr} did not take account of the evolution of the PQ field during inflation which can be relevant to the suppression of axion isocurvature perturbations.
Moreover, their analysis of the thermal mass for the PQ field was based on the high-temperature expansion of the thermal potential.
However, multi-loop effects are actually comparable to the leading-order effect, and they can significantly affect the shape of the thermal effective potential.
Therefore, in order to identify the parameter space that realizes the PQ symmetry non-restoration precisely, it is essential to properly resum the multi-loop effects, as recently demonstrated in the context of the electroweak symmetry non-restoration~\cite{Baldes:2018nel,Glioti:2018roy,Meade:2018saz,Carena:2021onl}.

The present paper pursues the idea of non-restoration of the PQ symmetry to address the domain wall problem and makes it in a complete form.
The QCD axion $a$ originates from a complex scalar field $\Phi$ that is charged under the $U(1)_{\rm PQ}$ symmetry.
Spontaneous breaking of $U(1)_{\rm PQ}$ is labeled by a non-vanishing vacuum expectation value (VEV) of the radial part $\phi = \sqrt{2}|\Phi|$, $v_{\rm PQ}\equiv\langle \phi \rangle \neq 0 $.
The axion decay constant is then defined as $f_a \equiv v_{\rm PQ}/ N_{\rm DW}$. 
We assume $f_a\ll T_{\rm R}$ and take the KSVZ axion framework~\cite{Kim:1979if,Shifman:1979if} for demonstration purposes.
In this framework, to achieve a high axion quality usually leads to $N_{\rm DW} > 1$~\cite{Lu:2023ayc} (see e.g. Refs.~\cite{Nakai:2021nyf,Qiu:2023los,Nakagawa:2023shi,Nakagawa:2024kcb} for recent papers on high-quality axion models with $N_{\rm DW}>1$), so that the domain wall problem arises.
To resolve the issue, we protect the PQ broken phase by introducing real singlet scalar(s) $s$. 
Then, a PQ scalar-singlet mixing interaction, $\lambda_{\phi s} \phi^2 s^2$ with $\lambda_{\phi s}<0$, gives a negative correction to the thermal mass of $\phi$, which makes $v_{\rm PQ}(T)>0$ for all $T\leq T_{\rm R}$.

A potential way out of the axion isocurvature problem can be seen in Eq.~\eqref{isocurvature}, as originally proposed by Linde~\cite{Linde:1991km} where if the PQ field has a much larger VEV during inflation, the axion fluctuation is significantly suppressed.
In fact, taking $f_{\rm inf} \simeq 0.1 M_{\rm Pl}$ and $f_a = 10^{12} \, \rm GeV$ in Eq.~\eqref{isocurvature}, the axion isocurvature bound becomes $H_{\rm inf} \lesssim 1.1\times 10^{14} \, \rm GeV$ under $\theta_i = 0.1$.
Such a situation is able to be achieved by introducing a (negative) Hubble-induced mass term for the PQ breaking field $\phi$. 
However, $\phi$ starts to evolve after inflation, which may trigger the parametrically enhanced production of the PQ field, leading to the non-thermal restoration of the PQ symmetry or the production of axion domain walls~\cite{Kasuya:1996ns,Kasuya:1997td,Kawasaki:2013iha,Harigaya:2015hha,Ema:2017krp,Kawasaki:2017kkr,Kawasaki:2018qwp}.
Refs.~\cite{Kasuya:1996ns,Ema:2017krp,Kawasaki:2017kkr,Kawasaki:2018qwp} argued that the shortcoming of the Linde's proposal can be avoided by considering a supersymmetric axion model with a flat direction.
In this case, the PQ symmetry is never restored by the parametric resonance effect~\cite{Ema:2017krp,Kawasaki:2017kkr}.~\footnote{Note that the fluctuations of the axion are produced via the parametric resonance. 
For the axion domain wall to be produced, the fluctuation must give an exactly flat distribution to the axion field at the QCD scale.
Although a detailed simulation is required for concluding the fate, we assume that the axion distribution does not become completely flat, resulting in no domain walls.}
In the present study, we consider a (non-supersymmetric) model with two PQ fields which make a flat direction, as in the case of the supersymmetric model, and evaluate the thermal potential, taking account of the multi-loop contributions, to find the parameter space that realizes non-restoration of the PQ symmetry.
%~\footnote{
%\red{Supersymmetry can be broken by thermal corrections~\cite{Das:1978rx}. Our purpose here is to demostrate that PQ nonrestoration can be realized in a simpler setup. Thus, we consider the nonsupersymmetric model.}}
Our model then lies outside the conventional classification into pre- or post-inflationary QCD axion models, while simultaneously addressing all cosmological problems that arise in both frameworks.

\section{Suppressing the isocurvature}

To suppress the isocurvature fluctuations, one possible idea is to consider a flat direction for the PQ scalar to evolve from a large value to today's one. 
We utilize the proposal of Refs.~\cite{Kasuya:1996ns,Ema:2017krp,Kawasaki:2017kkr,Kawasaki:2018qwp} in a form without supersymmetry. 
Consider two complex scalar fields $\Phi_\pm$, which are charged oppositely under $U(1)_{\rm PQ}$. The tree-level potential is given by
\begin{equation}
    V_0  = \lambda \left| \Phi_+ \Phi_- - v^2 \right|^2 + m_+^2 \left| \Phi_+ \right|^2 + m_-^2 \left| \Phi_- \right|^2 \;,
    \label{tree}
\end{equation}
where $\lambda = \mathcal O(1)$ and $m_\pm$ are the mass parameters. 
Here we take $m_\pm = \mu$ for simplicity. 
In the limit of $m_\pm \rightarrow 0$,
the potential admits an exact flat direction,
\begin{equation}
\Phi_+\Phi_- = v^2\;.
\label{eq:flat}
\end{equation}
The tree-level potential has a vacuum that is located at $|\langle \Phi_+ \rangle| = |\langle \Phi_- \rangle| = v$ for $m_\pm = \mu$.
The vacuum breaks the global $U(1)_{\rm PQ}$ symmetry, and the axion emerges.
To single out the axion degree of freedom, we first let
\begin{equation}
    \Phi_+ = \frac{\Phi_1 + \Phi_2}{\sqrt{2}} \;,\quad \Phi_- = \frac{\Phi_1^* - \Phi_2^*}{\sqrt{2}}\;.
\end{equation}
In the case of $m_+\neq m_-$, the field redefinition should be altered accordingly.
Next, we perform a nonlinear decomposition as
\begin{equation}
    \Phi_1 = \frac{\phi}{\sqrt{2}} e^{i a/v_{\rm PQ}}\;,\quad \Phi_2 = \frac{\xi + i \eta }{\sqrt{2}} e^{i a/v_{\rm PQ}}\;,
\end{equation}
where $a$ is the axion, and $\{\phi,\xi,\eta\}$ are real fields. 
One can check that $V_0$ does not contain $a$ dependence.
The kinetic terms are
\begin{align}
    \mathcal L &\supset  \left| \partial \Phi_+\right|^2 + \left| \partial \Phi_- \right|^2  =  \left| \partial\Phi_1\right|^2 + \left| \partial\Phi_2 \right|^2 \\
    & = \frac{1}{2} \left(\partial \phi \right)^2 + \frac{1}{2} \left(\partial \xi \right)^2  + \frac{1}{2} \left(\partial \eta \right)^2 \nonumber \\
    &\qquad + \frac{1}{2} \frac{\phi^2 + \xi^2 + \eta^2}{v_{\rm PQ}^2} \left(\partial a \right)^2 
    + \dfrac{\xi \partial \eta - \eta \partial \xi }{v_{\rm PQ}}\partial a  \;.\nonumber
\end{align}
To canonically normalize the axion, the PQ scale is 
\begin{equation}
    v_{\rm PQ} = \sqrt{\langle \phi \rangle^2 + \langle \xi \rangle^2 + \langle \eta \rangle^2}\;.
    \label{eq:PQ_scale}
\end{equation}
In this model, $\phi$, $\xi$ and $\eta$ are all PQ fields, whose VEV determines the PQ scale.
Note that in the present work we ignore the kinetic mixing between $a$ and $\xi,\eta$  when calculating thermal corrections as it is suppressed by the PQ scale.

As discussed in the Introduction, we consider the scenario that a PQ scalar has a large field value during inflation to suppress the isocurvature perturbation.
$\Phi_\pm$ are stabilized at some values located at the flat direction~\eqref{eq:flat} due to the (negative) Hubble induced mass.
Without loss of generality, we assume $|\Phi_+|_{\rm inf}\simeq M_{\rm Pl}$ and $|\Phi_-|_{\rm inf} \simeq v^2/M_{\rm Pl}$, which gives $(v_{\rm PQ})_{\rm inf} \simeq M_{\rm Pl}$, and the isocurvature fluctuation of the axion is suppressed according to Eq.~\eqref{isocurvature}. ~\footnote{In our setup, during inflation, we have $\eta \ll \xi$ and $\xi \sim \phi$. So kinetic mixing between $\xi$ and $a$ reaches the maximum. The current decomposition is suitable for evaluating the thermal potential. While it is better adopt usual polar coordinate decomposition when calculating the cosmic evolution of fluctuations~\cite{Kasuya:1996ns}.}
The existence of the flat direction makes the subsequent evolution of PQ fields under control.
The non-supersymmetric potential in general includes additional quartic terms like $\Phi_+^2\Phi_-^2$ or $\Phi_+\Phi_-|\Phi_\pm|^2$. However, they will destroy the existence of flat direction, which shall create domain walls during the oscillating phase of the inflaton.

\section{Thermal potential}

During the inflation, there is no thermal bath. 
At the end of inflation, inflaton oscillates around its potential minimum before decay), all PQ fields starts to evolve towards their vacuum minimum.
There is no (negligible) thermal corrections during these two stages.

After the inflaton decay, the Universe goes through the reheating epoch, within which the SM thermal plasma is formed with a reheating temperature $T_{\rm R}$.
We consider high reheating temperature, which indicates short or instantaneous inflaton decay.
This indicates that all PQ fields start with large field values.
The existence of the flat direction (even in presence of the thermal corrections) suppresses the parametric production of the axion during the PQ fields stabilization process.
Thanks to the large mass perpendicular to the flat direction, the deviated fluctuations are suppressed during inflation.
Therefore, one only need to consider the potential vacuum solution.
To prevent the PQ symmetry restoration due to the temperature correction, we introduce $N_s$ scalars that couple to the PQ fields as $\lambda_{\phi s} s^2 \left(\Phi_+ \Phi_- +{\rm c.c.}\right)$, with $\lambda_{\phi s}<0$.
They form a $N_s$ representation of $O(N_s)$ symmetry and $s^2$ is the singlet.
In addition to the mixing terms, the tree-level potential~\eqref{tree} is explicitly rewritten by
\begin{align}
V_0 & = -\frac{1}{2}\mu_\phi^2 \phi^2 +  \frac{1}{2}\mu_\xi^2\xi^2 + \frac{1}{2}\mu_\eta^2 \eta^2 \nonumber\\
& \qquad  + \frac{\lambda}{8} \left( \eta^2 \phi^2 -\phi^2 \xi^2 +   \eta^2 \xi^2 \right) + \frac{\lambda}{16} \left(\phi^4 + \xi^4 + \eta^4 \right) \nonumber\\
&\qquad + \frac{\lambda_{\phi s}}{2} s^2 \phi^2 - \frac{\lambda_{\phi s}}{2} s^2 \left(\xi^2 + \eta^2 \right) \nonumber\\
& \qquad + y_j \phi \bar{\psi}_j \psi_j + \frac{\mu_s^2}{2}s^2+\frac{\lambda_s}{4}s^4\;,
\label{eq:tree_potential}
\end{align}
where $\mu_s$ and $\lambda_s$ are the mass parameter and quartic coupling for $s$, and $\mu_\phi^2 = \lambda v^2 - \mu^2$ and $\mu_\xi^2 =\mu_\eta^2 = \lambda v^2 + \mu^2$.~\footnote{One can also write down the coupling like $\lambda_\pm s^2 |\Phi_\pm|^2$, which introduces more quartic couplings between $s$ and PQ fields. Here for simplicity, we include only $\lambda_{\phi s}$ in the main text. }
Here we assume that $\mu^2< \lambda v^2$ such that $\mu_{\phi,\xi,\eta}^2>0$.
The constant contribution is neglected here.
The $\psi_j$ and $\bar \psi_j$ are heavy KSVZ quarks and $y_j$ is the Yukawa coupling.~\footnote{The mixing coupling of $\psi$ with the SM particles is assumed not to be small, e.g $H q_L \psi$.
Even if they are produced copiously from the dynamics of the PQ scalars, they can decay into thermal bath immediately, so that the dynamics of the PQ scalars is not back-reacted.
The detailed analysis was studied in Ref.~\cite{Moroi:2013tea}.} 
We assume that they are diagonalized for simplicity and $j = 1,2,\cdots, N_\psi$. 
The Yukawa coupling originates from terms like $\left(\Phi_+ + \Phi_-^*\right)\overline \psi \psi$, which is simplified to $\phi\overline\psi \psi$.
All dimensionless couplings except for $\lambda_{\phi s}$ are positive.
The potential is bounded from below as long as $|\lambda_{\phi s} | < \sqrt{\lambda_s \lambda}/2$.
The quartic term $s^4$ is understood as $(s\cdot s)^2$ which is the only quartic term allowed by $O(N_s)$ symmetry.~\footnote{The $O(N_s)$ symmetry for $N_s>1$ allows the mixing in the quartic term $s^4 =c_{klmn}  s_k s_l s_m s_n $ where $k,l, m, n$ run over $1,\cdots N_s$. 
Here $c_{klmn}$ is the Clebsch-Gordan coefficient projecting the 4th-order tensor product of $O(N_s)$ fundamental representations into the trivial representation.
In general, $c_{klmn}$ can take any linear combination of the following forms:
$$
c_{klmn}^{(1)} =\delta_{kl}\delta_{mn},\quad c_{klmn}^{(2)} =\delta_{km}\delta_{ln}, \quad c_{klmn}^{(3)} =\delta_{kn}\delta_{lm}.
$$
This can be verified from the tensor product decomposition,
\begin{align*}
\ytableausetup{smalltableaux} 
\ydiagram{1}\otimes \ydiagram{1}\otimes \ydiagram{1}\otimes \ydiagram{1}  
=& \left({\bf 1} \oplus\ydiagram{1,1}\oplus\ydiagram{2} \right)\otimes \left({\bf 1} \oplus\ydiagram{1,1}\oplus\ydiagram{2} \right) \\
=& 3  \times{\bf 1} \oplus 6 \, \ydiagram{1,1} \oplus 6 \, \ydiagram{2} \oplus \ydiagram{4} \\
&\quad \oplus 3 \, \ydiagram{3,1} \oplus 2 \, \ydiagram{2,2} \oplus 3 \, \ydiagram{2,1,1} \oplus \ydiagram{1,1,1,1} \;,
\end{align*}
where trivial representations $\bf 1$ represent (one of) ways to take trace over the tensor product, and all the other Young diagrams are understood as traceless representations.
Since quartic terms $s_k s_l s_m s_n $ are identical under exchanges of indices, we take $c_{klmn}= \delta_{kl}\delta_{mn}$ hereafter.}
The tree-level potential admits a vacuum which is $\phi = 2\mu_\phi/\sqrt{\lambda}$ and $\xi = \eta = s = 0$.
At the vacuum, $\phi, \xi$ and $\eta$ obtain squared masses $2\mu_\phi^2, \mu_\xi^2-\mu_\phi^2$ and $\mu_\eta^2+\mu_\phi^2$ respectively, so that $\xi$ is identified as the saxion.

The heavy quarks $\psi_j$ couple to the SM through QCD, which indicates that PQ fields were also once in the thermal equilibrium with the SM plasma.
Consider the finite temperature corrections $V^\beta$. 
In the high-temperature limit, they contribute to the effective temperature mass term of $\phi$, which is
\begin{equation}
    V^\beta \supset \frac{1}{4} \left( \sum_j y_j^2 + \frac{1}{8}\lambda + \frac{1}{6} N_s \lambda_{\phi s} \right) \phi^{2}T^{2}\;.
    \label{eq:high_T_V_1}
\end{equation}
In the absence of $s$, $\lambda_{\phi s} \to 0$, the thermal corrections from both the fermions $\psi$ and $\phi$ itself would lead to a positive effective quadratic term, which overcomes the negative $\mu^2$-term in $V_0$~\eqref{eq:tree_potential} and restores the PQ symmetry.
However, with a proper $|\lambda_{\phi s}|$, this effective quadratic term can stay negative, preventing the symmetry restoration.

\begin{figure}[!t]
    \centering
    \includegraphics[width=8.5cm]{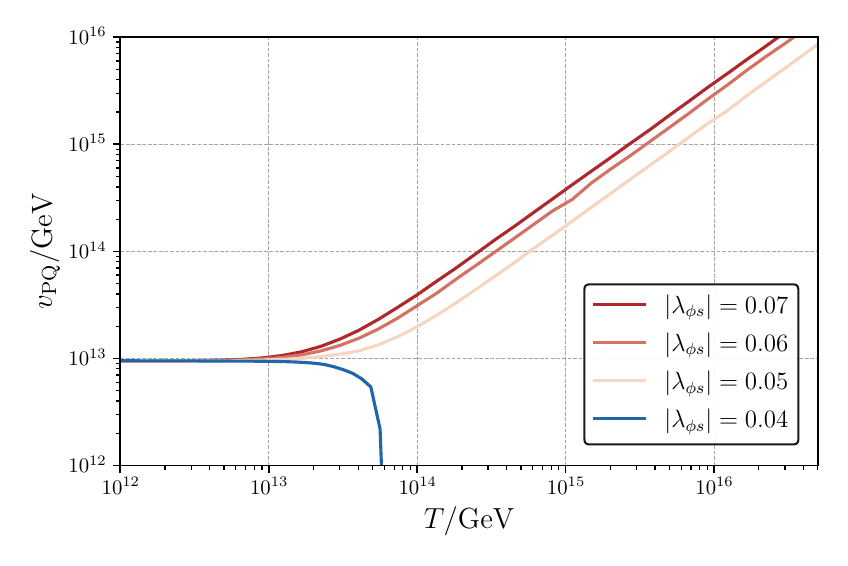}
    \caption{The PQ scale $v_{\rm PQ}$ against temperature $T$.
    Here we take $f_a = 10^{12}$~GeV, $\lambda =0.03$, $m_\psi = 10^{11}$~GeV, $\mu_s = 4\times 10^{12}$~GeV, $\mu_\xi = \mu_\eta = 2 \times 10^{12}$~GeV, $\lambda_s = 0.8$, $N_s = 1$ and $N_{\rm DW} = N_\psi = 10$.}
    \label{fig:vT}
\end{figure}

The effective potential with thermal corrections can be loop-expanded as
\begin{equation}
    V_{\rm eff}(\phi, \xi, \eta, s, T) = V_0 + V_1 + \cdots\;,
\end{equation}
where $V_0$~\eqref{eq:tree_potential} is the tree-level potential. 
Here we adopt the imaginary-time (Matsubara) formalism~\cite{Matsubara:1955ws}.
When the temperature goes higher, corrections from long wavelength modes become more significant. 
Then, one should include such effects by resuming multi-loop diagrams, which manifest themselves as thermal mass corrections in the propagators. 
The leading-order correction is 
\begin{align}
    V_1(T)  & = \sum_{i=\phi,\xi,\eta,s,\psi_j} \frac{n_i T}{2} \sum_{n = -\infty}^\infty \int \frac{d^3 \vec k}{(2\pi)^3}  \label{eq:1-loop thermal}\\
    & \quad \times \log \left[ \vec k^2 + \omega_n^2 + m_i^2(\phi,\xi,\eta,s) +\Pi_i(T) \right]\;,  \nonumber
\end{align}
where $(\omega_n, \vec k)$ is the Euclidean loop 4-momentum, $\omega_n$ are the Matsubara frequencies. 
The numbers of degrees of freedom are given by $n_{\{\phi,\xi,\eta,s,\psi_j\}} = \{1,1,1,N_s, -12\}$. 
The background-dependent masses are obtained by expanding the $V_0$~\eqref{eq:tree_potential} around the background scalar values, which are
\begin{subequations}
\begin{align}
    m_\phi^2 & = -\mu_\phi^2 + \lambda_{\phi s}  s^2 + \frac{3\lambda}{4} \phi^2 - \frac{\lambda}{4} \xi^2 + \frac{\lambda}{4} \eta^2 \ , \\
    m_\xi^2 & = \mu_\xi^2 - \lambda_{\phi s}  s^2 - \frac{\lambda}{4} \phi^2 + \frac{3\lambda}{4} \xi^2 + \frac{\lambda}{4} \eta^2 \ , \\
    m_\eta^2 & = \mu_\eta^2 - \lambda_{\phi s}  s^2 + \frac{\lambda}{4} \phi^2 + \frac{\lambda}{4} \xi^2 + \frac{3\lambda}{4} \eta^2 \ , \\[0.5ex]
    m_s^2  & = \mu_s^2 + 3\lambda_s s^2 + \lambda_{\phi s} \phi^2  - \lambda_{\phi s}  \left( \xi^2 + \eta^2 \right)  \ , \label{eq:ms2}\\[1ex]
    m_{\psi_j}^2 & = y_j^2 \phi^2\;.
\end{align}
\end{subequations}
The finite temperature mass shifts (Debye masses) are given by calculating the self-energy in the high temperature limit~\cite{Carrington:1991hz,Parwani:1991gq},
\begin{subequations}
\begin{align}
    \Pi_\phi & = \left(\sum_j \frac{y_j^2}{2} + \frac{\lambda}{16} + N_s \frac{\lambda_{\phi s}}{12}\right)T^2 \ , \\ 
    \Pi_\xi & = \left( \frac{\lambda}{16} - N_s \frac{\lambda_{\phi s}}{12} \right) T^2 \ , \\[0.5ex]
    \Pi_\eta & = \left( \frac{5\lambda}{48} - N_s \frac{\lambda_{\phi s}}{12} \right) T^2 \ , \\
    \Pi_s & = \left((N_s + 2)\frac{\lambda_s}{12} - \frac{\lambda_{\phi s}}{12}\right)T^2 \;,
\end{align}\label{eq:Pi_s}
\end{subequations}
which are valid for $N_s \ge 1$. 
The coefficient $(N_s+2)\lambda_s/12$ comes from symmetry factors of the one-loop Feynman diagrams concerning $s^4$. 
Consider the self-energy for a specific $s_i$ ($i = 1,2,\cdots, N_s$). If it is the same $s_i$ running in the internal loop, the symmetry coefficient is $3\lambda_s/12$. 
Meanwhile, if a different $s_{j\neq i}$ running in the loop, it gives a factor of $\lambda_s/12$.
Summing up all diagrams gives a total coefficient $(N_s+2)\lambda_s/12$ for a specific $s_i$.
There is no IR divergence for fermion in the high temperature limit, i.e. $\Pi_\psi =0$.
The correction $V_1$ can be split into three parts, namely a zero-temperature part, a one-loop $T$-dependent part, and a ``daisy" contribution from resuming multiple loop effects in the high temperature limit, $V_1= V_1^{(0)} + V_1^\beta + V_{\rm daisy}$, which are
\begin{align}
    V_1^{(0)} & = \sum_{i = \phi,\xi,\eta,s,\psi_j} \frac{n_i m_i^4 }{64\pi^2}\left(\log \frac{m_i^2}{m_{0i}^2} - \frac{3}{2} \right) \;, \nonumber \\ 
    V_1^\beta & = \sum_{i = \phi,\xi,\eta,s}  \frac{n_i T^4}{2\pi^2} J_b\left(\frac{m_i^2}{T^2}\right) + \sum_{j = 1}^{N_\psi} \frac{n_{\psi_j}T^4}{2\pi^2} J_f \left( \frac{m_i^2}{T^2} \right) \;, \nonumber \\
    V_{\rm daisy} & = \sum_{i=\phi,\xi,\eta,s}\frac{n_i T}{12\pi}\left[m_i^3 - \left( m_i^2 + \Pi_i \right)^{3/2}\right] \;,
\end{align}
where $m_{0i}$ are masses evaluated at the vacuum solution,
\begin{equation}
    \langle \phi \rangle = f_a N_{\rm DW}\;,\quad \langle \xi \rangle = \langle \eta \rangle = \langle s \rangle = 0\;.
    \label{eq:vacuum_solution}
\end{equation}
Choosing a different $m_{0i}$ is equivalent to changing the renormalization scheme.
We fix the renormalization scheme so that all parameters here are measured physical parameters.
The functions $J_{b/f}$ are defined as
\begin{equation}
    J_{b/f}(x) = \int_0^\infty dk k^2 \log \left[1 \mp \exp{\left(-\sqrt{k^2 + x}\right)} \right] \;.
\end{equation}
The effective potential could be complex at some field values. We take the real part to determine the stabilization of the potential. The imaginary part can be interpreted as tunneling effects, and is usually small compared to the real part~\cite{Delaunay:2007wb}.

The temperature dependence of the PQ scale $v_{\rm PQ}(T)$ in Eq.~\eqref{eq:PQ_scale} is solved by stabilizing the effective potential at its global minimum, which satisfies
\begin{equation}
    \frac{\partial V_{\rm eff}}{\partial \phi} = \frac{\partial V_{\rm eff}}{\partial \xi}=\frac{\partial V_{\rm eff}}{\partial \eta}=\frac{\partial V_{\rm eff}}{\partial s} = 0\;.
\end{equation}
We choose the parameter such that $\langle \xi \rangle = \langle \eta \rangle = \langle s \rangle =0$ stay the same as the vacuum against high temperatures.
By randomly sampling the initial field values and using the steepest descent method, we have checked that the vacuum solution here is indeed the global minimum, and there is no other metastable minimums along the flat direction.
As shown in Fig.~\ref{fig:vT}, one can make sure that $v_{\rm PQ}(T)\neq 0$ for any high temperature with a proper $|\lambda_{\phi s}|$.

\begin{figure}
    \centering
    \includegraphics[width=8.5cm]{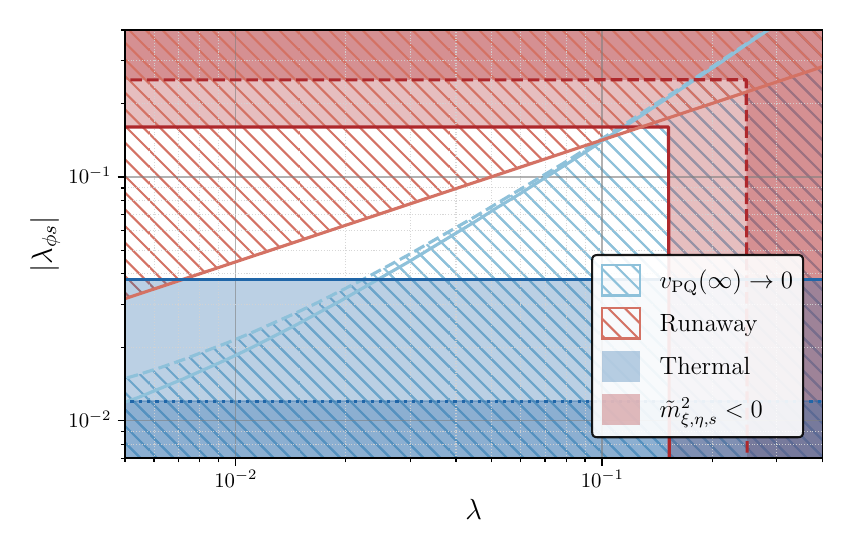}
    \caption{The parameter space for $|\lambda_{\phi s}|$ and $\lambda$.
    The shaded regions are excluded by different aspects. 
    The runaway region is where $|\lambda_{\phi s} | > \sqrt{\lambda_s \lambda}/2$.
    The $v_{\rm PQ}(\infty) \to 0$ indicates an insufficient amount of $|\lambda_{\phi s}|$ to keep the broken phase against high temperature according to Eq.~\eqref{eq:high_T_V_1}.
    The $\tilde m_{\xi,\eta,s}^2<0$ region indicates the instability of expected vacuum solution.
    Here $\tilde m_i^2$ is the second derivative of the zero temperature potential $V_0 + V_1^{(0)}$ evaluated at Eq.~\eqref{eq:vacuum_solution}.
    Thermal effects will further stabilize them as indicated by their thermal mass corrections~\eqref{eq:Pi_s}.
    Here we take $m_\psi = 10^{11}$~GeV, $\mu_\xi = \mu_\eta = 2\times 10^{12}$~GeV, $\mu_s = 4\times 10^{10}$~GeV, $\lambda_s=0.8$, $N_s =1$ and $N_{\rm DW} = N_\psi = 10$ as a benchmark.
    The solid (dashed) lines are limits under $f_a = 10^{12} \, (8\times 10^{11}) \, \rm GeV$.
    The thermal bounds (blue shaded region) are where $s$ does not have thermal contact with PQ fields; the solid (dotted) line is obtained under reheating temperature $T_{\rm R} = 10^{15} \, (10^{14}) \, \rm GeV$. Here we take the Hubble parameter during radiation-dominated epoch as $H\simeq 3.4 \times T^2/M_{\rm Pl}$.
    }
    \label{fig:par_space}
\end{figure}

Here for the purpose of demonstration of a viable parameter space, we treat all the heavy quarks have the same mass $m_\psi$, and the number of which~\footnote{Here $N_\psi$ is the number of Dirac fermions.} determines the DW number, $N_{\psi} = N_{\rm DW} = 10$. 
Then, the Yukawa coupling is given by $y_j =  m_\psi / \sqrt{2} f_a N_{\rm DW}$.
Fig.~\ref{fig:par_space} shows the allowed parameter space.
The mass parameter $\mu_\phi$ can be determined by stabilizing the zero-temperature potential $V_0 + V_1^{(0)}$ at Eq.~\eqref{eq:vacuum_solution}.
The mass parameter $\mu_{\xi,\eta, s}$ should be large enough to protect the stability of the vacuum solution~\eqref{eq:vacuum_solution}, both against the quantum and thermal corrections.
We show that even the simplest scenario, $N_s = 1$, can give a viable parameter space. A larger $N_s$ will further release more parameters (a smaller $|\lambda_{\phi s}|$ is allowed for a larger $N_s$ as indicated by Eq.~\eqref{eq:high_T_V_1}).
As we take a smaller $f_a$, one needs a larger $|\lambda_{\phi s}|$ to keep $U(1)_{\rm PQ}$ broken against high temperature. Meanwhile, the pressure of stabilizing Eq.~\eqref{eq:vacuum_solution} is smaller.
In order for $s$ to have thermal corrections to the PQ fields, it has to enter the thermal equilibrium via $ss\to\phi\phi \, (\xi\xi, \eta\eta)$, whose interaction rates can be estimated as $\Gamma \sim |\lambda_{\phi s}|^2 T$ for high temperature. 
During the radiation-dominated epoch, the Hubble parameter is given by $H \sim T^2/M_{\rm Pl}$.
Only when $\Gamma> H$, $s$ can be in thermal contact with the plasma.
Therefore, for $T> T_{\rm crit} \sim |\lambda_{\phi s}|^2 M_{\rm Pl}$, $s$ decouples from the thermal bath. As long as $T_{\rm crit} > T_{\rm R}$, the reheating temperature, $s$ can help protect broken PQ symmetry to avoid the domain wall problem.

\section{Summary and discussion}

In this work, we have explored a cosmological scenario in which the PQ symmetry remains broken throughout the entire thermal history of the Universe. This setup eliminates the formation of axion strings and domain walls without relying on late-time symmetry breaking or additional bias terms. 
We have shown that, contrary to conventional expectations, the inclusion of scalar interactions—even if there is only one extra scalar—can naturally prevent symmetry restoration at high temperatures. 
By considering two complex scalars as PQ fields, one can create a flat direction, which allows a PQ field to remain stabilized at a large value during inflation, effectively suppressing axion isocurvature perturbations and avoiding the parametric resonance production of topological defects after reheating. 
Our detailed analysis of the finite-temperature effective potential reveals a viable parameter space for the PQ symmetry non-restoration. This provides a minimal and robust framework that simultaneously addresses the domain wall and isocurvature problems while allowing arbitrarily high reheating temperatures, making it fully compatible with high-scale baryogenesis mechanisms such as thermal leptogenesis.

Let us emphasize that it was a non-trivial issue to accommodate both the PQ symmetry non-restoration and Linde's approach to the isocurvature problem. 
The PQ symmetry can be restored due to not only thermal effects but also non-thermal particle production which results from the dynamics of a PQ scalar during and after inflation.
There is an approach to consider the sextet potential which reduces the oscillation amplitude of the PQ field at the balance point with the quartic term~\cite{Harigaya:2015hha}.
However, we have found that this approach does not work for the PQ symmetry non-restoration successfully.
This is because the fluctuations produced via the parameteric resonance effect  may restore the PQ symmetry for a large quartic coupling constant.
Taking a smaller quartic coupling leads to a smaller $|\lambda_{\phi s}|$ to get the potential bounded from below, and thus, the sector for $s$ cannot be thermalized. 
This fact motivates us to consider the current setup.

If supersymmetry is considered, the flat direction~\eqref{eq:flat} is guaranteed without fine tuning. Then one has to include superpartners when evaluating the thermal potentials.
Supersymmetry can be broken by thermal corrections~\cite{Das:1978rx}. Thus, it is also possible to have a sizable effect from $\lambda_{\phi s}$ to protect the PQ broken phase against high temperature. However, the dynamics becomes complicated.
Our purpose here is to demonstrate that PQ nonrestoration can be realized in a simpler setup. Thus, we consider the nonsupersymmetric model, and leave the full supersymmetric realization for future work.

\noindent {\it Note added.}---
As our project was being completed, we were aware of an overlapping work~\cite{Lee:2025blt}.

\begin{acknowledgments}

We thank Fuminobu Takahashi for useful discussions and coordination. LW thanks Jiang Zhu for valuable discussions.
YN is supported by Natural Science Foundation of Shanghai. The work of Y.~-C. Qiu is supported by GRF Grants No.~11302824 and No.~11310925 and CityUHK Grants No.~9610645 and No.~7020130

\end{acknowledgments}

\bibliography{reference}

\end{document}